\documentclass[epjCONF]{svjour}
\usepackage{graphicx}
\usepackage[varg]{txfonts} 
\usepackage[latin1]{inputenc}
\def\bsigma{\mbox{\boldmath$\sigma$}}
\def\btau{\mbox{\boldmath$\tau$}}
\session-title{%
19$^{\textnormal{\footnotesize th}}$ International %
IUPAP Conference on Few-Body Problems in Physics%
}
\begin{document}
\title{%
Treatment of Two Nucleons in Three Dimensions
}%
\author{%
I. Fachruddin\inst{1}\fnmsep\thanks{\email{imamf@fisika.ui.ac.id}}
\and %
Ch.~Elster\inst{2}
\and %
J.~Golak\inst{3}
\and %
R.~Skibi\'nski\inst{3}
\and %
W.~Gl\"ockle\inst{4}
\and %
H.~Wita{\l}a\inst{3}
}
\institute{%
Departemen Fisika, Universitas Indonesia, Depok 16424, Indonesia
\and %
Institute of Nuclear and Particle Physics,
Department of Physics and Astronomy, Ohio University, Athens, OH 45701, USA
\and %
M. Smoluchowski Institute of Physics, Jagiellonian
University, PL-30059 Krak\'ow, Poland
\and %
Institut f\"ur Theoretische Physik II,
Ruhr-Universit\"at Bochum, D-44780 Bochum, Germany
}
\abstract{
We extend a new treatment proposed for two-nucleon (2N) and three-nucleon (3N) bound states to 2N scattering.
This technique takes momentum vectors as variables, thus, avoiding partial wave decomposition, and handles
spin operators analytically. We apply the general operator structure of a nucleon-nucleon (NN) potential to
the NN T-matrix, which becomes a sum of six terms, each term being scalar products of spin operators and momentum
vectors multiplied with  scalar functions of vector momenta. Inserting this expansions of the NN force and
T-matrix into the Lippmann-Schwinger equation allows to remove the spin dependence by taking traces
and
yields a set of six coupled equations for the scalar functions found in the expansion of the T-matrix.
} 
\maketitle
%
%
%
\section{Introduction}

In Ref.~\cite{2N3N} a new formulation for the 2N and 3N bound states in three dimensions has been proposed.
In this technique momentum vectors are taken as variables, avoiding a traditional partial-wave decomposition. In addition
spin operators occurring as scalar products of spin and momentum vectors - shortly called spin-momentum
operators - are evaluated analytically by means of trace operations. In this approach a NN force is
employed using its most general operator structure, i.e. as  sum of 6 spin-momentum operators multiplied with
scalar functions of momenta. A spin-\linebreak momentum operator representation is used as well for
the 2N and 3N bound states, as in Refs.~\cite{deut} and \cite{oph3}, respectively.

We extend the technique developed in Ref.~\cite{2N3N} to NN scattering. This would be an alternative to other
three-\linebreak dimensional approach formulated in a momentum-helicity basis~\cite{NN3D}. In addition we
introduce a new set of spin-\linebreak momentum operators different from the one used in \linebreak
Ref.~\cite{2N3N}. We find one of the spin-momentum operators in Ref.~\cite{2N3N} violates time reversal 
and, therefore, has to be multiplied with a time-reversal violating scalar function. Here we prefer to work with 
operators, which are also invariant with respect to time reversal.   
The idea is to apply the general operator structure not only to the NN force but also to
the NN T-matrix. The goal is then to find the scalar functions in the expansion of the T-matrix into the
spin-momentum operators. First, we insert the spin-momentum operators expansions of the NN interaction and
T-matrix into the Lippmann-Schwinger equation. Next by analytical evaluation we remove the spin dependence
yielding finally a set of coupled equations for the scalar functions of the T-matrix. Finally we connect
the T-matrix to the anti-symmetrized scattering amplitude parameterized by the \linebreak Wolfenstein parameters.

\section{Formulation}\label{formulation}

\subsection{The general operator structure of NN potential}

The general operator structure of NN potential reads
\begin{equation}
V^ { t m_t}({\bf p'}, {\bf p}) = \sum_{j=1}^ 6 v_j^ { t m_t} ({\bf p'}, {\bf p}) \;
w_j({\bsigma}_1,{\bsigma}_2, {\bf p'}, {\bf p}) , \label{vinw}
\end{equation}
with $V^ { t m_t}({\bf p'}, {\bf p})$ being the NN potential projected on the NN total isospin states $ \mid t m_t\rangle$ as
\begin{equation}
V^ { t m_t}({\bf p'}, {\bf p}) = \langle  t m_t \mid V({\bf p'}, {\bf p}) \mid t m_t\rangle .
\end{equation}
The scalar functions $v_j^ { t m_t} ({\bf p'}, {\bf p})$ depend only on the vector momenta.
the
$w_j({\bsigma}_1,{\bsigma}_2, {\bf p'}, {\bf p})$ are a set of spin-momentum operators,
\begin{eqnarray}
w_1 ( {\bsigma}_1,{\bsigma}_2, {\bf p'}, {\bf p})&  = &  1\cr
w_2 ( {\bsigma}_1,{\bsigma}_2, {\bf p'}, {\bf p})&  = & {\bsigma}_1 \cdot {\bsigma}_2\cr
w_3 ( {\bsigma}_1,{\bsigma}_2, {\bf p'}, {\bf p)}&  = & i \; ( {\bsigma}_1
+ {\bsigma}_2 ) \cdot ( {\bf p} \times {\bf p'})\cr
w_4 ( {\bsigma}_1,{\bsigma}_2, {\bf p'}, {\bf p})&  = & {\bsigma}_1
\cdot ( {\bf p} \times {\bf p'}) \; {\bsigma}_2 \cdot ( {\bf p} \times {\bf p'})\cr
w_5 ( {\bsigma}_1,{\bsigma}_2, {\bf p'}, {\bf p})&  = & {\bsigma}_1
\cdot  ({\bf p'} + {\bf p}) \; {\bsigma}_2 \cdot  ({\bf p'} + {\bf p})\cr
w_6 ( {\bsigma}_1,{\bsigma}_2, {\bf p'}, {\bf p})&  = & {\bsigma}_1
\cdot ( {\bf p'} - {\bf p}) \; {\bsigma}_2 \cdot  ( {\bf p'} - {\bf p}) ,
\end{eqnarray}
which is time-reversal invariant. As an example a leading order (LO) chiral NN potential is given as~\cite{evgeny.report} \pagebreak
\begin{eqnarray}
V_{LO}({\bf p'}, {\bf p}) & = & -\frac{1}{(2\pi)^3}  \frac{g_A^2}{4 F_\pi^2} \frac{ w_6 ( {\bsigma}_1,{\bsigma}_2, {\bf p'}, {\bf p})}{( {\bf p'} - {\bf p})^2 +M_\pi^2} {\btau_1} \cdot {\btau_2}  \cr
&& + \frac{C_S}{(2\pi)^3}w_1 ( {\bsigma}_1,{\bsigma}_2, {\bf p'}, {\bf p}) \cr
&& + \frac{C_T}{(2\pi)^3}w_2 ( {\bsigma}_1,{\bsigma}_2, {\bf p'}, {\bf p}) .
\end{eqnarray}

\subsection{The deuteron}

We briefly describe the formulation for the deuteron. The deuteron has total spin $1$ and isospin $0$.
In spin-momentum operator representation the deuteron state is given as~\cite{deut}
\begin{equation}
\Psi_{m_d}({\bf p}) = \langle {\bf p} | \Psi_{m_d}\rangle = \sum_{ k=1}^2 \phi_k(p) \;  b_k( {\bsigma}_1, {\bsigma}_2, {\bf p}) | 1 m_d \rangle , \label{deutwf}
\end{equation}
where $| 1 m_d \rangle$ is the total-spin state with magnetic quantum number $m_d$, $\phi_k(p)$ scalar functions depending on the magnitude of momenta only, and $b_k( {\bsigma}_1, {\bsigma}_2, {\bf p})$ spin-momentum operators given as
\begin{eqnarray}
b_1( {\bsigma}_1, {\bsigma}_2, {\bf p}) & = & 1 \cr
b_2( {\bsigma}_1, {\bsigma}_2, {\bf p}) & = & {\bsigma}_1 \cdot {\bf p} \; {\bsigma}_2 \cdot {\bf p} - \frac{1}{3} p^2 .
\end{eqnarray}
The scalar functions $\phi_k(p)$ are connected to the standard partial-wave projected deuteron
 s-wave $\psi_0(p)$ and d-wave $\psi_2(p)$ by~\cite{deut}
\begin{eqnarray}
\psi_0 (p) & = & \phi_1 (p) \cr
\psi_2 (p) & = & \frac{4 p^2} { 3 \sqrt{2}} \, \phi_2 (p)  .
\end{eqnarray}
Inserting $\Psi_{m_d}({\bf p})$ of Eq.~(\ref{deutwf}) and $V^ { t m_t}({\bf p'}, {\bf p})$ of Eq.~(\ref{vinw}) into the Schr\"odinger equation  for the deuteron in integral form,
\begin{equation}
\Psi_{m_d}({\bf p}) = \frac{1}{ E_d -\frac{p^2}{m}} \int d^3 p' V^ {00}({\bf p}, {\bf p'}) \Psi_{m_d}({\bf p'}),
\end{equation}
yields
\begin{eqnarray}
\lefteqn{\sum_{ k=1}^2 \phi_k(p) \;  b_k( {\bsigma}_1, {\bsigma}_2, {\bf p}) | 1 m_d \rangle} && \cr
&& \qquad = \frac{1}{ E_d -\frac{p^2}{m}} \int d^3 p' \sum_{j=1}^ 6 v_j^ {00} ({\bf p}, {\bf p'}) \;
w_j({\bsigma}_1,{\bsigma}_2, {\bf p}, {\bf p'}) \cr
&& \qquad \quad \sum_{ k'=1}^2 \phi_{k'}(p) \;  b_{k'}( {\bsigma}_1, {\bsigma}_2, {\bf p'}) | 1 m_d \rangle . \label{dschr}
\end{eqnarray}

To remove the spin dependence from Eq.~(\ref{dschr}) we project Eq.~(\ref{dschr}) on $\langle 1 m_d | b_i( {\bsigma}_1, {\bsigma}_2, {\bf p})$ from the left and sum up over $m_d$. We obtain
\begin{eqnarray}
\sum_{ k=1}^2 A^ d_{i k}(p) \phi_{k} (p) & = & \frac{1}{ E_d-\frac{p^2}{m}}\int d^3 p' \sum_{j=1}^6 v_j^ {00}( {\bf p},{\bf p'}) \cr
&& \sum_{k'=1}^2 B^ d_{ijk'} ( {\bf p},{\bf p'})\phi_{k'}(p') ,
\end{eqnarray}
which is a set of two coupled equations for $\phi_{k} (p)$, with $A^ d_{i k}(p)$ and $B^ d_{ijk'} ( {\bf p},{\bf p'})$ being defined as
\begin{equation}
A^ d_{i k} ( p) \equiv \sum_{ m_d = -1}^{1 } \langle 1 m_d|
b_i( {\bsigma}_1, {\bsigma}_2, {\bf p})
 b_k( {\bsigma}_1, {\bsigma}_2, {\bf p})| 1 m_d \rangle 
\end{equation}
\begin{eqnarray}
B^ d_{ijk'} ( {\bf p},{\bf p'}) & \equiv & \sum_{ m_d = -1}^{1 } \langle 1 m_d|
b_i( {\bsigma}_1, {\bsigma}_2, {\bf p}) w_j( {\bsigma}_1, {\bsigma}_2, {\bf p}, {\bf p'}) \cr
&& \qquad b_{k'}( {\bsigma}_1, {\bsigma}_2, {\bf p'})| 1 m_d \rangle .
\end{eqnarray}
The functions $A^ d_{i k}(p)$ and $B^ d_{ijk'} ({\bf p},{\bf p'})$ are scalar functions of the vectors ${\bf p}$ and ${\bf p'}$, and need to be calculated only once.
As example we have e.g.
\begin{eqnarray}
A^d_{11} (p) & = &  3\cr
A^d_{22} (p) & = &  \frac{8}{3} p^4 \cr
B^d_{141} ({\bf p},{\bf p}^{\prime}) & = & ({\bf p}\times {\bf p'})^2 \cr
B^d_{151} ({\bf p},{\bf p}^{\prime}) & = & ({\bf p'}+ {\bf p})^2 \cr
B^d_{161} ({\bf p},{\bf p}^{\prime}) & = & ({\bf p'}- {\bf p})^2 .
\end{eqnarray}

\subsection{The NN scattering}

The operator structure given in Eq.~(\ref{vinw}) for NN force can also be applied to the NN T-matrix as
\begin{equation}
T^ { t m_t}({\bf p'}, {\bf p}) = \sum_{j=1}^ 6 t_j^ { t m_t} ({\bf p'}, {\bf p}) \;
w_j({\bsigma}_1,{\bsigma}_2, {\bf p'}, {\bf p}) , \label{tinw}
\end{equation}
with $t_j^ { t m_t} ({\bf p'}, {\bf p})$ being the scalar functions to be found. Inserting both
the expansion in Eqs.~(\ref{vinw}) and (\ref{tinw}) into the Lippmann-Schwinger equation,
\begin{eqnarray}
T^ { t m_t}({\bf p'}, {\bf p}) & = & V^ { t m_t}({\bf p'}, {\bf p}) \cr
&& + 2 \mu \lim_{\epsilon \rightarrow 0} \int d{\bf p''} \frac{V^ { t m_t}({\bf p'}, {\bf p''}) T^ { t m_t}({\bf p''}, {\bf p})}{p^2 + i\epsilon -p''^2} , \qquad
\end{eqnarray}
where $\mu$ is the reduced mass of the NN system, leads to
\begin{eqnarray}
\lefteqn{\sum_{k=1}^ 6 t_k^ { t m_t} ({\bf p'}, {\bf p}) \; w_k({\bsigma}_1,{\bsigma}_2, {\bf p'}, {\bf p})} && \cr
&& \qquad = \sum_{k=1}^ 6 v_k^ { t m_t} ({\bf p'}, {\bf p}) \; w_k({\bsigma}_1,{\bsigma}_2, {\bf p'}, {\bf p}) \cr
&& \qquad \quad + 2 \mu \lim_{\epsilon \rightarrow 0} \int d{\bf p''} \frac{1}{p^2 + i\epsilon -p''^2} \cr
&& \qquad \quad \sum_{j=1}^ 6 v_j^ { t m_t} ({\bf p'}, {\bf p''}) \; w_j({\bsigma}_1,{\bsigma}_2, {\bf p'}, {\bf p''}) \cr
&& \qquad \quad \sum_{k'=1}^ 6 t_{k'}^ { t m_t} ({\bf p''}, {\bf p}) \; w_{k'}({\bsigma}_1,{\bsigma}_2, {\bf p''}, {\bf p}) . \label{ls}
\end{eqnarray}
We remove the spin dependence from Eq.~(\ref{ls}) by multiplying from the left with $w_i({\bsigma}_1,{\bsigma}_2, {\bf p'}, {\bf p})$ and perform the trace. 
This leads to \pagebreak
\begin{eqnarray}
\lefteqn{\sum_{k = 1}^6 A_{ik}({\bf p'}, {\bf p}) t_k^{tm_t}({\bf p'}, {\bf p}) } && \cr
&& \qquad = \sum_{k = 1}^6 A_{ik}({\bf p'}, {\bf p}) v_k^{tm_t}({\bf p'}, {\bf p}) \cr
&& \qquad \quad + 2 \mu \lim_{\epsilon \rightarrow 0} \sum_{j,k' = 1}^6 \int d{\bf p''}
   \frac{1} {p^2 + i\epsilon - p''^2} \cr
&& \qquad \quad B_{ijk'}({\bf p'}, {\bf p''}, {\bf p})
                v_j^{tm_t}({\bf p'}, {\bf p''}) t_{k'}^{tm_t}({\bf p''}, {\bf p}) , \label{nnls}
\end{eqnarray}
where $A_{ik}({\bf p'}, {\bf p})$ and $B_{ijk'}({\bf p'}, {\bf p''}, {\bf p})$ are defined as
\begin{eqnarray}
A_{ik}({\bf p'}, {\bf p}) & \equiv & Tr \{w_i({\bsigma}_1,{\bsigma}_2, {\bf p'}, {\bf p}) \cr
&& \quad \; w_k({\bsigma}_1,{\bsigma}_2, {\bf p'}, {\bf p})\} \\
B_{ijk'}({\bf p'}, {\bf p''}, {\bf p}) & \equiv & Tr \{w_i({\bsigma}_1,{\bsigma}_2, {\bf p'}, {\bf p}) w_j({\bsigma}_1,{\bsigma}_2, {\bf p'}, {\bf p''}) \cr
&& \quad \; w_{k'}({\bsigma}_1,{\bsigma}_2, {\bf p''}, {\bf p})\}
\end{eqnarray}
Again the functions
 $A_{ik}({\bf p'}, {\bf p})$ and $B_{ijk'}({\bf p'}, {\bf p''}, {\bf p})$ are scalar functions of the momenta ${\bf p}$ and ${\bf p'}$, and need to be evaluated only once.
As example we show here,
\begin{eqnarray}
A_{24}({\bf p}', {\bf p}) & = & A_{42}({\bf p}', {\bf p}) = 4 ({\bf p} \times {\bf p}')^2 \cr
A_{56}({\bf p}', {\bf p}) & = & A_{65}({\bf p}', {\bf p}) = 4 (p'^2 - p^2)^2  \cr
B_{122}({\bf p}', {\bf p}'', {\bf p}) & = & B_{212}({\bf p}', {\bf p}'', {\bf p}) = B_{221}({\bf p}', {\bf p}'', {\bf p}) = 12 \cr
B_{124}({\bf p}', {\bf p}'', {\bf p}) & = & B_{214}({\bf p}', {\bf p}'', {\bf p}) = 4 ({\bf p} \times {\bf p}'')^2 \cr
B_{144}({\bf p}', {\bf p}'', {\bf p}) & = & 4 \{({\bf p}'' \times {\bf p}') \cdot ({\bf p} \times {\bf p}'')\}^2
\end{eqnarray}
Equation (\ref{nnls}) is a set of six coupled equations for $t_k^{tm_t}({\bf p'}, {\bf p})$, which can e.g. be solved
as a system of linear equations.

The NN scattering observables can be calculated from the anti-symmetrized scattering amplitude $M^ {t m_t}_{ m_1' m_2', m_1 m_2}({\bf p'},{\bf p})$, \linebreak which is defined as
\begin{eqnarray}
\lefteqn{M^ {t m_t}_{ m_1' m_2', m_1 m_2}({\bf p'},{\bf p})} && \cr
&& \; \equiv \langle t m_t | \langle m_1' m_2'| \langle {\bf p'}| M ( 1 - P_{12}) | {\bf p}\rangle | m_1 m_2\rangle | t m_t\rangle
\end{eqnarray}
and can be parameterized by the Wolfenstein parameters $a^{t m_t}({\bf p'},{\bf p})$, $c^{t m_t}({\bf p'},{\bf p})$, $m^{t m_t}({\bf p'},{\bf p})$, $g^{t m_t}({\bf p'},{\bf p})$, $h^{t m_t}({\bf p'},{\bf p})$ \linebreak as~\cite{book}
\begin{eqnarray}
\lefteqn{M^ {t m_t}_{ m_1' m_2', m_1 m_2}({\bf p'},{\bf p})} && \cr
&& \quad = a^ {tm_t}({\bf p'},{\bf p}) \; \langle m_1' m_2'| w_1( {\bsigma}_1, {\bsigma}_2, {\bf p'},{\bf p}) | m_1 m_2\rangle \cr
&& \qquad -  i \frac{ c^ {tm_t}({\bf p'},{\bf p})}{ | {\bf p} \times {\bf p'}|} \;
 \langle m_1' m_2'|w_3 ({\bsigma}_1, {\bsigma}_2, {\bf p'}, {\bf p})| m_1 m_2\rangle\cr
&& \qquad +  \frac{ m^ {tm_t}({\bf p'},{\bf p})}{ | {\bf p} \times {\bf p'}|^ 2}
 \langle m_1' m_2'| w_4( {\bsigma}_1, {\bsigma}_2, {\bf p'}, {\bf p}) | m_1 m_2\rangle \cr
&& \qquad +  \frac{ g^ {tm_t}({\bf p'},{\bf p})+h^ {tm_t}({\bf p'},{\bf p})}{ ( {\bf p} + {\bf p'})^ 2} \cr
&& \qquad \quad \langle m_1' m_2'|  w_5 ( {\bsigma}_1, {\bsigma}_2, {\bf p'},{\bf p}) | m_1 m_2\rangle \cr
&& \qquad +  \frac{ g^ {tm_t}({\bf p'},{\bf p})-h^ {tm_t}({\bf p'},{\bf p})}{ ( {\bf p} - {\bf p'})^ 2} \cr
&& \qquad \quad \langle m_1' m_2'| w_6( {\bsigma}_1, {\bsigma}_2, {\bf p'},{\bf p}) | m_1 m_2\rangle . \quad \label{wpar}
\end{eqnarray}
Thus, finally we connect the scattering amplitude or similarly the Wolfenstein parameters to the scalar function \linebreak $t_j^ { t m_t} ({\bf p'}, {\bf p})$. This can be accomplished by means of Eq.~(\ref{wpar}) and the relation
between the M- and T-matrix given as
\begin{equation}
M = -\mu (2\pi)^2 T .
\end{equation}
We obtain
\begin{eqnarray}
a^ {tm_t} & = &  t_1 + (-)^ t \; \Big[ \frac{1}{2} \tilde t_1 + \frac{3}{2} \tilde t_2 \cr
&& +   \frac{1}{2} p^ 4 ( 1-x^ 2)  \tilde t_4 + p^ 2 ( 1-x)  \tilde t_5 + p^ 2 (
1+x) \tilde t_6\Big] \cr
c^ {tm_t} & = &  i p^ 2 \sqrt{1 - x^ 2}  \left( t_3 - (-)^ t \tilde t_3 \right) \cr
m^ {tm_t} & = &  t_2 + p^ 4 ( 1 - x^ 2)  t_4
 +   (-)^ t \; \Big[ \frac{1}{2} \tilde t_1 - \frac{1}{2} \tilde t_2 \cr
&& + \frac{1}{2} p^ 4 ( 1- x^ 2) \tilde t_4 -   p^ 2 (1-x)\tilde t_5 - p^ 2 ( 1+x) \tilde t_6 \Big]\cr
g^ {tm_t} & = &  t_2 + p^ 2 ( 1+x)  t_5  + p^ 2( 1-x)  t_6 \cr
&& +  (-)^ t \; \Big[  \frac{1}{2} \tilde t_1 - \frac{1}{2} \tilde t_2
- \frac{1}{2} p^ 4 ( 1-x^ 2) \tilde t_4 \Big]\cr
h^ {tm_t} & = &  p^ 2(1+x)  t_5 - p^ 2(1-x) t_6 \cr
&& +  (-)^ t \; \Big[- p^ 2(1-x)  \tilde t_5 + p^ 2(1+x)  \tilde t_6 \Big] . \label{wpart},
\end{eqnarray}
where $x = {\bf \hat p'} \cdot {\bf \hat p}$
Note that in Eq.~(\ref{wpart}) we drop ${\bf p}$ and ${\bf p'}$ for simplicity and apply the following notation,
\begin{eqnarray}
t_j & \equiv & t_j^ {t m_t} ( {\bf p'}, {\bf p}) \cr
\tilde t_j &\equiv & t_j^ {t m_t} ( {\bf p'}, -{\bf p}) .
\end{eqnarray}

\section{Summary}

We propose a new technique to calculate the 2N system as function of momentum vectors, i.e.
 without employing a partial wave decomposition. The technique
is  useful especially in energy regions of hundreds of MeV or when considering the NN t-matrix as input to a three-body calculation. Based on scalar interactions, the scattering of three-bosons has been successfully carried out up to the GeV regime, formulating the Faddeev equations as functions of vector momenta~\cite{Liu:2004tv}. The formulation of NN scattering presented here is an important
step on the way of performing realistic three-body scattering calculations at higher energies.

 Based on the general operator
structure of the NN interaction we derive the formulation in a spin-momentum operator representation. Here
the NN potential, the T-matrix, and the deuteron state are expanded in a set of scalar products
of spin operators and momentum vectors. We derive a set of two coupled equations for the deuteron wave function
components, which are connected to the standard partial wave projected wave function s- and d-wave in a simple
manner. In case of the NN scattering we obtain a set of six coupled equations for the scalar functions defining
the NN T-matrix, and therefore,  the scattering amplitude in the Wolfenstein representation.

\end{document}